\documentclass[12pt,a4paper]{article}
\begin{document}

\centerline{\large \bf Comment on:}
\centerline{\large \bf ``The Supersymmetric Ward-Takahashi Identity in 1-Loop}
\centerline{\large \bf Lattice Perturbation Theory. I. General Procedure''}
\centerline{\large \bf by A. Feo; hep-lat/0305020}
\vspace{5mm}

T. Galla$\rm^a$, G. M\"unster$\rm^b$ and A. Vladikas$\rm^c$

(a) Department of Physics,
University of Oxford, Theoretical Physics,\\
1 Keble Road, Oxford OX1 3NP, UK

(b) Institut f\"ur Theoretische Physik,
Universit\"at M\"unster,\\
Wilhelm-Klemm-Str. 9, D-48149 M\"unster, Germany

(c) INFN, Sezione di Roma 2, c/o Dipartimento di Fisica,\\
Universit\'a di Roma ``Tor Vergata'',\\
Via della Ricerca Scientifica 1, I-00133 Rome, Italy
\vspace{5mm}

\centerline{\bf Abstract}
We comment on several issues concerning both the physics and authorship
of article\\
hep-lat/0305020.
\vspace{5mm}

In a recent paper A.~Feo studies the supersymmetric Ward identities for N=1
supersymmetric Yang-Mills theory on a lattice, in the framework of lattice
perturbation theory. Her article, being the first in a series of two, deals
with the general procedure and the methodological foundations of the
calculation. We have some critical comments on the contents of this paper.

1. The lattice SUSY Ward identity is considered in terms of the bare
lattice operators, see eqs.~(3.2), (3.8). From these equations eq.~(3.11)
does not follow, contrary to what is claimed. A multiplicative
renormalization of contact terms, Fadeev-Popov term and gauge fixing term
with renormalisation constants $Z_{CT}$, $Z_{GF}$ and $Z_{FP}$ could
result from some of the operators $B_j$ appearing in (3.8). In (3.11) the
renormalization constants as well as the sum over all operators $B_j$
appear. Therefore these contributions are double-counted.

2. After eq.~(3.13), it is claimed that $Z_{S}$, $Z_{T}$ and $Z_{\chi}$ are
power subtraction coefficients of the operator $X_S / a$ and therefore, based
on ref.~[36], their independence from a renormalisation scale is derived as a
corollary. It is certainly wrong to claim that $Z_\chi$ is a power
subtraction coefficient (this is true of $\tilde m$ in eq.~(3.6)). The
operator $\chi(x)$ is multiplicatively renormalisable. Thus, $Z_\chi$ has a
logarithmic divergence $\ln(a\mu)$. Only in this way can the RG invariance of
the product $(m_0 - \tilde m) \chi$ (implicit in the  Ward identities of the
paper) be true. Since on the other hand, $Z_S$ and $Z_T$ are shown to be
finite after eq.~(3.13), we do not understand the following statement in the
conclusions:~``We observe that, at least at 1-loop order in perturbation
theory, $Z_T$ is finite. This result may have some theoretical implications
which we are currently investigating.'' In our view, the problem is not the
finiteness of $Z_S$ and $Z_T$, but whether these coefficients are all that is
required to renormalise the SUSY current. In this respect the situation with
supersymmetric Ward identities is not analogous to the one dealt in
ref.~[36], concerning chiral Ward identities and the normalisation $Z_A$ of
the axial current.

3. The Faddeev-Popov term is already $O(g_{0}^2)$ at tree-level, as remarked
after (4.30). Therefore it contributes to the Ward identity to order
$g_{0}^2$ and should appear in (4.30), where it is missing. In other words,
the claim $Z^{(0)}_{FP} =0$, made right after eq.~(4.28) is unjustified.

4. Before (4.34) the definition
$\Delta \equiv {\cal O} \nabla_\mu S_\mu(x) +
\frac{ \delta {\cal O}}{\delta \bar \xi(x)}|_{\xi = 0} +
{\cal O}  \, \frac{\delta S_{GF}}{\delta \bar \xi(x)}|_{\xi = 0} +
{\cal O} B_i $ is not consistent with (4.31), (4.32), as it contains the
terms ${\cal O} B_i$, whose second order contributions do not appear in
(4.31), (4.32). The contributions from ${\cal O} B_i$ should not be included
in (4.34).

5. The strategy for calculating the renormalisation constants without
evaluating the one-loop contribution of the complicated term $X_{S}$
requires knowledge of the list of mixing operators $B_{j}$ appearing in
(3.8). This implies an analysis of possible BRS-exact operators of
dimension smaller than $\frac{11}{2}$. This analysis has not been carried out
and it is not clear whether $B_{0}$, $B_{1}$, $B_{2}$ and $B_{3}$ form a
complete set. For example, $B_4 = \frac{2}{g} (\partial_{\rho} A_{\mu})
\gamma_{\mu} \partial_{\rho} \lambda$, being independent of the other
operators $B_j$, should have been considered at the general level of
eq.~(3.10), although its contribution vanishes at the specific kinematical
regime of eq.~(4.33).

6. The work presented in A.~Feo's paper is part of a joint project within
the framework of the DESY-M\"unster-Roma collaboration. All involved
researchers and students had agreed on joint publications of their results.
The main contribution of A.~Feo is the detailed implementation, programming
and numerical evaluation of the various perturbative contributions. This
considerable work, which we appreciate, is announced for publication in the
sequel paper. The general procedure outlined in Feo's present paper, however,
grew out of discussions and internal written memoranda by the three of us.
In Refs.~[37,38,40] these sources are acknowledged as private communications
and internal notes. None of us, however, has authorized their external use.
Moreover, any definitive publication was postponed until the open questions
(exposed in points 2.\ and 5.\ above) were resolved. Thus, so far we have
limited ourselves to the joint presentation of partial results in past
lattice workshops (see hep-lat/0110113, hep-lat/0011030).

We regret that last year A.~Feo decided to cease collaborating with us. We
express our total disapproval of this unauthorized publication of the
collaboration's internal documents on unfinished work. We declare that we will
refrain from any other public discussion or further comment on this matter,
be it on the present preprint archive or elsewhere.

\end{document}